\def\hmpc{\ifmmode{h^{-1}\,\hbox{Mpc}}\else{$h^{-1}$\thinspace Mpc}\fi}
\def\hkpc{\ifmmode{h^{-1}\,\hbox{kpc}}\else{$h^{-1}$\thinspace kpc}\fi}

\def\kms{\ifmmode{\,\hbox{km}\,s^{-1}}\else {\rm\,km\,s$^{-1}$}\fi}

\def\mpc{{\rm\,Mpc}}
\def\msun{{\rm\,M_\odot}}
\def\lsun{{\rm\,L_\odot}}

\documentstyle{rspublic_mod}
\input{epsf.sty}

\newcommand{\Mpccube}{\mbox{\,h$^{-3}$\,Mpc$^3$}}
\newcommand{\gs}
           {\mathrel{\hbox{\rlap{\hbox{\lower4pt\hbox{$\sim$}}}\hbox{$>$}}}}
\newcommand{\ls}
           {\mathrel{\hbox{\rlap{\hbox{\lower4pt\hbox{$\sim$}}}\hbox{$<$}}}}

\begin{document}

\title{The CNOC2 Field Galaxy Redshift Survey}
\shorttitle{The CNOC2 Redshift Survey}
\author{R.G.~Carlberg$^{1,5,6}$, H.K.C.~Yee$^{1,6}$, S.L.~Morris$^{2,6}$, 
	H.~Lin$^{1,6}$, M.~Sawicki$^{1,6}$, G.~Wirth$^{3,6}$,
        D.~Patton$^{3,6}$, C.W.~Shepherd$^1$, E.~Ellingson$^{4,6}$,
        D.~Schade$^{2,6}$, C.J.~Pritchet$^3$, \& F.D.A.~Hartwick$^3$}
\shortauthor{The CNOC2 Collaboration}
\affiliation{$^1$Department of Astronomy, University
        of Toronto, $^2$Dominion Astrophysical Observatory, Herzberg
        Institute of Astrophysics, National Research Council of
        Canada, $^3$Department of Physics \& Astronomy, University of
        Victoria, $^4$Center for Astrophysics \& Space Astronomy,
        University of Colorado, $^5$Observatories of the Carnegie
        Institution of Washington, $^6$Visiting Astronomers,
        Canada--France--Hawaii Telescope, which is operated by the
        National Research Council of Canada, le Centre National de
        Recherche Scientifique, and the University of Hawaii.}

\maketitle

\abstract 
The CNOC2 field galaxy redshift survey (hereafter CNOC2) is designed
to provide measurements of the evolution of galaxies and their
clustering over the redshift range 0 to 0.7. The sample is spread over
four sky patches with a total area of about 1.5 square degrees. Here
we report preliminary results based on two of the sky patches, and the
redshift range of 0.15 to 0.55.  We find that galaxy evolution can be
statistically described as nearly pure luminosity evolution of early
and intermediate SED types, and nearly pure density evolution of the
late SED types. The correlation of blue galaxies relative to red
galaxies is similar on large scales but drops by a factor of three on
scales less than about 0.3\hmpc, approximately the mean scale of
virialization.  There is a clear, but small, 60\%, change in
clustering with 1.4 mag of luminosity. To minimize these population
effects in our measurement of clustering evolution, we choose galaxies
with $M_r^{k,e}\le -20$ mag as a population whose members are most
likely to be conserved with redshift.  Remarkably, the evolution of
the clustered density in proper co-ordinates at $r
\ls 10$\hmpc, $\rho_{gg}\propto r_0^\gamma(1+z)^3$, is best described
as a ``de-clustering'', $\propto (1+z)^{0.6\pm0.4}$. Or equivalently,
there is a weak growth of clustering in co-moving co-ordinates,
$x_0\propto (1+z)^{-0.3\pm0.2}$.  This conclusion is supported by the
pairwise peculiar velocities which rise slightly, but not
significantly, into the past.  The Cosmic Virial Theorem applied to
the CNOC2 data gives $Q\Omega_M/b=0.11\pm0.04$, where $Q$ is the three
point correlation parameter and $b$ the bias. Similarly, galaxy groups
have a virial mass-to-light ratio (evolution corrected) of $\langle
M_{virial}/L_R^{k,e} \rangle\simeq 215h \lsun/\msun$, or
$\Omega_M=0.15\pm0.05$. 
\endabstract

\section{The motivation and design of the survey}

\begin{figure}
\centering
\vspace*{5pt}
\parbox{\textwidth}{\epsfysize=7truecm\epsfbox{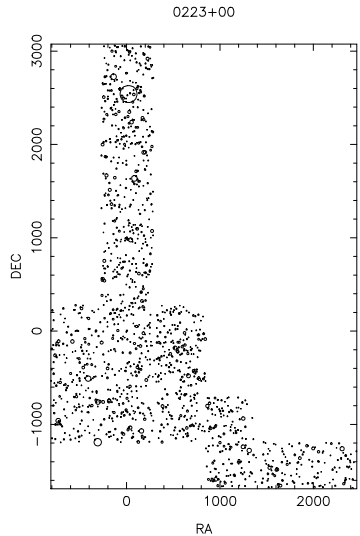}}
\vspace*{5pt}
\caption{The distribution on the sky, in seconds of arc, of the galaxies
        with redshifts and $m_R\le 21.7$ mag in 19 of the 20 fields in
        the 0223+00 patch.}
\label{fig:patch}
\end{figure}

The best test of theories of structure evolution is to observe the
evolution of both galaxies and their clustering.  The Canadian Network
for Observational Cosmology (CNOC) field galaxy redshift survey is
designed to investigate clustering dynamics and its relation to galaxy
evolution on scales smaller than approximately 20\hmpc\ over the $0\le
z \le 0.7$ range. To meet these goals requires a ``CfA-class'' survey
(\cite{dp}) which contains roughly $10^4$ galaxies in $10^6\Mpccube$
to allow subsampling and to provide minimal coverage of a
representative set of clustering environments. For observational
convenience the survey is distributed over four patches, each
nominally containing 20 Multi-Object Spectrograph, MOS, (\cite{mos})
fields of approximately $9^\prime\times8^\prime$ on the sky. The
layout (minus one field) is shown in Figure~\ref{fig:patch}.

\begin{figure}
\centering
\vspace*{5pt}
\parbox{\textwidth}{\epsfysize=8truecm\epsfbox{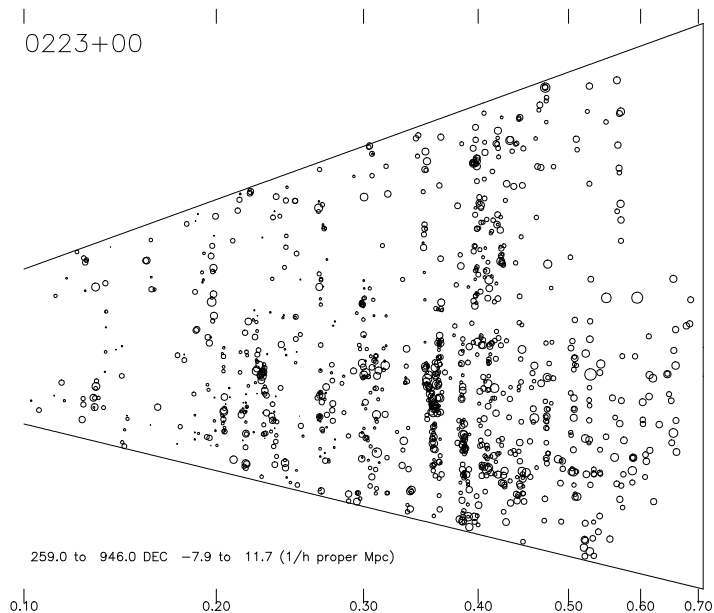}}
\vspace*{5pt}
\caption{The line-of-sight distance versus the transverse 
        distance in the Dec direction from the field centre, in proper
        co-ordinates, in the 0223+00 patch.}
\label{fig:pie}
\end{figure}

The fields are observed in the UBgRI filters, with the R
(Kron-Cousins) magnitudes (measured as total magnitudes, \cite{ppp})
being used to define the survey.  The photometric sample extends to
about $m_R=24$ mag, with comparable depths in the other filters,
except for U and I which extend to 23 mag.  The spectroscopic sample
is drawn from a ``mountain'' version of the photometric sample with a
nominal limit of $m_R=21.5$ mag. Each field is observed with two
spectroscopic masks, the ``bright'' mask extending to $m_R\simeq 20$
mag and the ``faint'' mask extending to the limiting magnitude. Slits
for additional fainter objects are placed in any otherwise unoccupied
areas.  Together the two masks largely eliminate the problem of slit
crowding and have the further benefit of increasing the efficiency of
the observations. The spectra are band limited with a filter extending
from 4400-6300\AA, which gives a statistically complete sample over
the $0.15\le z \le 0.55$ range, with emission line galaxies visible
over $0 \le z \le 0.7$. The average completeness of the resulting
redshift sample relative to the photometric sample is about 50\%. The
success rate for obtaining redshifts from spectra is about 85\% after
accounting for the fraction of objects expected to be outside our
passband. The failures are mainly due to poor seeing, poor
transparency and objects in the corners of the MOS.  The redshift
distribution for one patch is shown in Figure~\ref{fig:pie}.  The
results below are derived assuming $H_0= 100 h \kms\mpc^{-1}$ and
$q_0=0.1$.

\section{Evolution of the Luminosity Function}

The availability of UBgRI photometry for our sample allows us to
classify CNOC2 galaxies by colour and to compute the luminosity
functions (LF) for different galaxy populations in a number of
different bandpasses.  In particular, we give details of our LF
methods and descriptions of CNOC2 LF evolution results in
Lin \et\ (1998a), and confront our LF, number count, and colour
distribution data against a variety of galaxy evolution models in
Lin \et\ (1998b).

\begin{figure}
\centering
\vspace*{5pt}
\parbox{\textwidth}{\epsfysize=10truecm\epsfbox{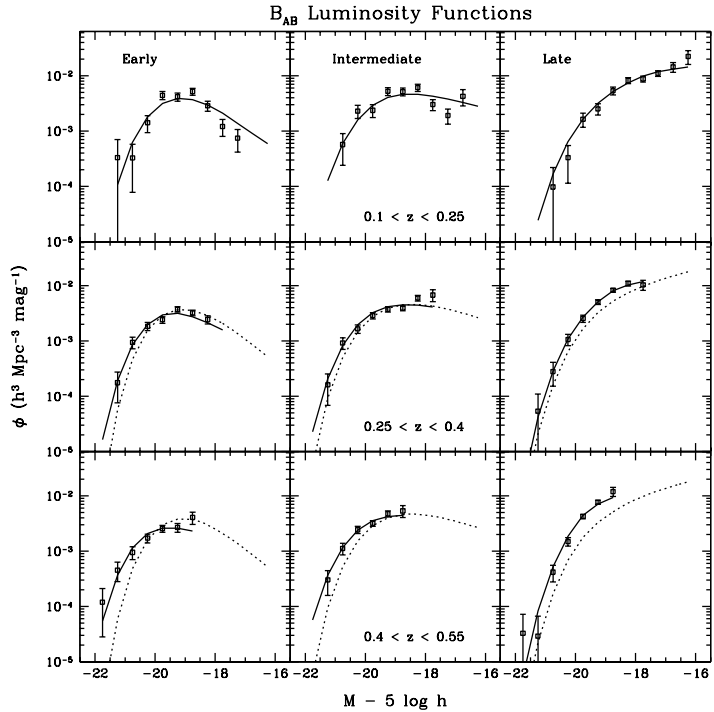}}
\vspace*{5pt}
\caption{
Evolution of the $B_{AB}$ luminosity functions (solid curves and
   points) for early-, intermediate-, and late-type CNOC2 galaxies in
   three redshift bins ($z$ increases from top to bottom).  Also shown
   are fiducial LF's (dotted curves) from the lowest-redshift bin for
   each galaxy type.  Results shown are for $q_0 = 0.1$.  }
\label{fig:p9B}
\end{figure}

\begin{figure}
\centering
\vspace*{5pt}
\parbox{\textwidth}{\epsfysize=7truecm\epsfbox{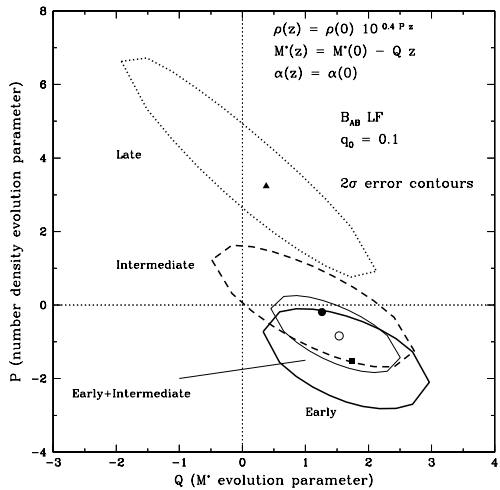}}
\vspace*{5pt}
\caption{
The two sigma confidence contours in $P$ (number density evolution
  parameter) versus $Q$ ($M^*$ evolution parameter) for the $B_{AB}$
  luminosity functions of early, intermediate, late, and
  early+intermediate CNOC2 samples.  The intersection of the
  horizontal and vertical dotted lines indicates no-evolution, $P = Q
  = 0$.  Results shown are for $q_0 = 0.1$.  }
\label{fig:pqB}
\end{figure}

In \cite{cnoc2lf1} we calculate LF parameters in the B$_{\rm AB}$, R,
and U bands for ``early,'' ``intermediate,'' and ``late'' CNOC2
galaxies, classified using fits of UBRI colours to the galaxy spectral
energy distributions (SED's) of \cite{cww}.  We present a description
of the LF evolution using the following convenient model,
\begin{eqnarray}
M^*(z) & = & M^*(0) - Q z \nonumber \\ 
\alpha(z) & = & \alpha(0) \nonumber \\ 
\rho(z) & = & \rho(0) 10^{0.4 P z} \nonumber \ ,
\end{eqnarray} 
where $M^*$ and $\alpha$ are the usual Schechter LF parameters, $\rho$
is the galaxy number density, and $P$ and $Q$ parameterize the rates
of number density evolution and luminosity evolution, respectively.
We plot our B$_{\rm AB}$ LF results in Figure~\ref{fig:p9B} and show
$2\sigma$ $P$-$Q$ error contours in Figure~\ref{fig:pqB}.  We find
that the faint-end slope of the LF is steeper for late-type galaxies
relative to early-type objects, consistent with previous LF studies at
both intermediate and low redshifts, ({\it e.~g.},
\cite{cfrslf,ldss,lcrslf}).  Moreover, the LF's of the early and
intermediate populations evolve differently from that of late-type
galaxies.  Specifically, we find that the LF's of early and
intermediate types show primarily positive luminosity evolution ($Q
\approx 1.5$) and only modest density evolution ($P \approx -1$),
while the late-type LF is best fit by strong positive number density
evolution ($P \approx 3$) and little luminosity evolution ($Q \approx
0.5$).  We also confirm the trend seen in previous smaller
intermediate-redshift samples that the luminosity density of late-type
galaxies increases strongly with redshift, but that the luminosity
density of early-type objects remains relatively constant with $z$.
These general conclusions hold for either $q_0 = 0.1$ (as in the LF
figures shown) or $q_0 = 0.5$.  Specific comparisons against the
Canada-France (\cite{cfrslf}) and Autofib (\cite{ldss}) redshift
surveys show general agreement among our LF evolution results,
although there remain some detailed discrepancies with respect to the
B-selected Autofib survey, which may be due to differences in galaxy
classification or sample selection methods.

In \cite{cnoc2lf1} we also compute SED type distributions, UBgRI
number counts, and various colour distributions for CNOC2 galaxies.
The number counts and colour distributions for {\em all} R $< 21.5$
CNOC2 galaxies (not just those with redshifts in our completeness
range $0.1 < z < 0.55$ used to compute the LF) are well matched once
we extrapolate our LF evolution models to $z \approx 0.75$, thus
providing additional checks on the validity of our LF evolution
results.  In addition, we have verified that various systematic
effects, specifically patch-to-patch variations, photometric errors,
surface brightness selection, redshift incompleteness, and apparent
magnitude incompleteness, do not significantly affect our results.

Subsequent papers on galaxy population evolution in CNOC2 will also
make use of the morphological and spectral information that will
become available for CNOC2 galaxies once the appropriate data are
fully reduced.  We are also in the process of deriving properly
calibrated photometric redshifts, which should provide another factor
of two increase in useful sample size for R $< 21.5$ galaxies.  Future
papers will further explore the issue of LF evolution using these even
larger CNOC2 galaxy data sets.

\section{Two Point Correlations}

\begin{figure}
\centering
\vspace*{5pt}
\parbox{\textwidth}{\epsfysize=8truecm\epsfbox{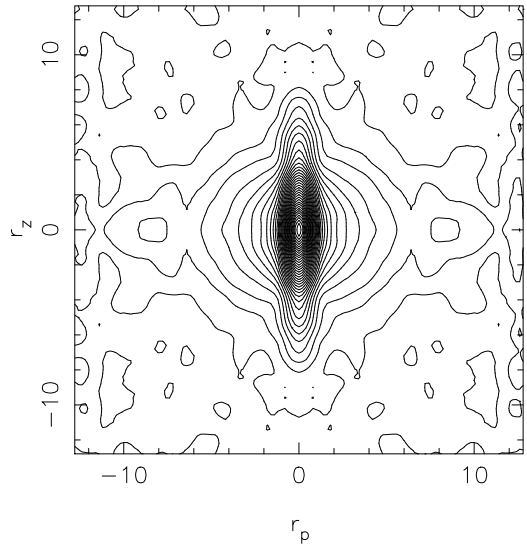}}
\vspace*{5pt}
\caption{The two dimensional correlation 
over the $0.15\le z \le 0.55$ range. The plot is symmetrized about the
centre and smoothed with a filter that increases with distance from
the origin. The apparent flattening of the contours at large distances
is not yet statistically significant.  }
\label{fig:2dcor}
\end{figure}

On scales where clustering is strongly nonlinear the two point
correlation function is a dynamically useful statistic, which is now
fairly well calibrated to initial conditions via n-body simulations
({\it e.~g.} \cite{virgo}). Our survey is sufficiently densely
sampled, with a velocity accuracy of 100 \kms\ or better, that we
normally derive all our results from the two dimensional correlation
function, $\xi(r_p,r_z)$, shown in Figure~\ref{fig:2dcor}, which
ratios excess pairs to the smooth background at projected separation
$r_p$ and redshift separation $r_z$. We estimate $\xi(r_p,r_z)$ with
the classical estimator $DD/DR -1$ (\cite{lss}), which has the
advantage of simplicity and speed. The $DD\cdot RR/(DR\cdot DR)-1$ and
$(DD-2DR+RR)/RR$ estimators lead to no significant changes of the
results presented here.

The real space correlation function can be derived from the projected
correlation function, $w_p(r_p)=\int \xi(\sqrt{r_p^2+r_z^2})\,dr_z$.
In the limit that the $r_z$ integral extends to infinity
$w_p(r_p)/r_p=
\Gamma(1/2)\Gamma((\gamma-1)/2)/\Gamma(\gamma/2)(r_0/r_p)^\gamma$.
The immediate complication with this measure of $\xi(r)$ is that when
applied to the data the sum needs to be truncated at some finite
$r_z$. The minimum $r_z$ is that required to sum over the random
pairwise velocities along the line of sight, say
$H(z)r_z\ge3\sigma_{12}$, or approximately 10\hmpc. At larger $r_z$
the sum converges roughly as $1-(r_0/r_z)$, that is, relatively
slowly. However, increasing the cutoff $r_z$ beyond about 50\hmpc\
increases the noise from large scale structure. The results here use
cutoff $r_z$ of 20 to 50 \hmpc, and fits to the resulting $w_p(r_p)$
use data with $r_p\le 10\hmpc$. We make no correction for correlation
beyond the cutoff $r_z$, which is generally less than 10\%, even for
the unrealistic pure power law correlation model.

\subsection{Luminosity and Colour Dependence of Correlations}

\begin{figure}
\centering
\vspace*{5pt}
\parbox{\textwidth}{\epsfysize=7truecm \epsfbox{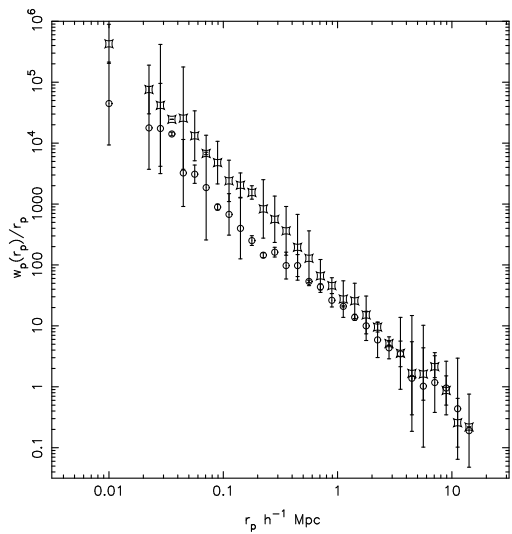}}
\vspace*{5pt}
\caption{Auto-correlations as a function of colour. The 
pincushions are for red galaxies, the circles are for the blue
galaxies. The blue galaxies appear to have a characteristic length for
correlation change of about 0.3\hmpc. The error flags are computed
from the difference of the two fields and greatly exceed the
$\sqrt{N}$ estimates.}
\label{fig:xi_br}
\end{figure}

\begin{figure}
\centering
\vspace*{5pt}
\parbox{\textwidth}{\epsfysize=7truecm\epsfbox{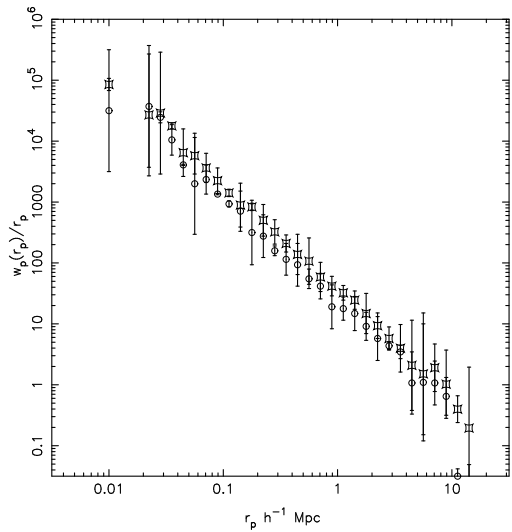}}
\vspace*{5pt}
\caption{Auto-correlations as a function of luminosity. The 
pincushions are for $M_R^{k,e}\le -20$ galaxies and the circles for
those between $-18.5$ and $-20$ mag.}
\label{fig:xi_lum}
\end{figure}

The auto-correlations of red and blue galaxies (of all luminosities)
are shown in Figure~\ref{fig:xi_br}. The sample is split approximately
in half at $(B-R)_0=1.25$ which divides the galaxies into those with
low and high star formation rate.  The resulting subsamples have
nearly identical mean redshifts, 0.35, but the red subsample has a
mean luminosity of $M_R=-20.5$ mag as compared to the blue $M_R=-19.8$
mag, so there will be some excess of red over blue as a result of
luminosity dependent correlation, see Figure~\ref{fig:xi_lum}. The
blue galaxy auto-correlation has a characteristic scale, roughly
0.3\hmpc, shortward of which they fall well below red galaxies.  A
similar trend in the Elliptical/Spiral ratio is reported in the APM
(\cite{apmz}) and CfA+SSRS2 survey (Marzke, private communication). A
comparison to the pairwise velocity dispersion, approximately 350\kms,
and the properties of groups found in these data (see below), suggests
that the ``blue break'' at 0.3\hmpc\ is the radius where field
galaxies statistically join a virialized region, inside of which star
formation is suppressed. The dynamics of virialization in a
hierarchical cosmology is to tidally remove the outer dark matter and
any imbedded baryons. The net effect is to eventually starve the
galaxy of infalling gas.

The issue of luminosity dependence of correlations is an important
test theories of primordial bias associated with density
``peaks''. The observational situation is unclear at the moment. The
APM survey finds about a factor of three increase in correlation for
about a 2 mag change in luminosity.  The LCRS (\cite{lcrs}) has no
significant luminosity dependence of clustering after allowance for
redshift differences, over a directly comparable range of
luminosities. A mild enhancement, 35\% in clustering amplitude, of the
clustering of high over low luminosity galaxies is seen in the
Perseus-Pisces catalogue (\cite{pp,guzzo}).

The dependence of clustering on luminosity is shown in
Figure~\ref{fig:xi_lum}. The sample is divided at $M_R=-20$, k
corrected and evolution corrected at an approximate rate of
$M_R(0)=M_R-Qz$, with $Q=1$.  The mean luminosities are $-19.35$ and
$-20.75$ mag. The $M_\ast$ of a Schechter fit to the luminosity
function is -20.3 mag. The mean redshifts of the two samples are 0.37
and 0.34. The fitted correlation lengths are 3.7\hmpc\ and 2.65\hmpc\,
both with $\gamma=1.7$, or a ratio of clustering amplitudes of high to
low luminosity of 1.67. These correlations are significantly larger
than estimated from the small sky area, somewhat lower luminosity,
samples available previously (\cite{cfrsxi,cnoc1xi,kkeck}).

Within the CNOC2 data the change of correlation amplitude with
luminosity is comparable in size with what is expected on the basis of
galaxies of increasing luminosities being correlated with dark matter
halos of increasing mass.  That is, very approximately, $\xi\propto
\Delta(M)^{-2} \propto M^{(n+3)/6}$, where $\Delta(M)$ is the mass
field variance for spheres containing mass $M$, which for a
perturbation spectrum, $P(k)\propto k^n$, is $\Delta\propto
M^{-(n+3)/6}$. For $n\simeq -2$ expected from CDM-like spectra on
galaxy scales, we then expect a correlation amplitude ratio of 1.58
for the factor of four difference in luminosity. That is, we attribute
the luminosity dependence of the correlation as reflecting an
underlying primordial difference in the correlations, roughly as
expected for galaxy formation in dark matter potential wells.

\section{Evolution of the Two Point Correlation Function}

At low redshift the evolution of galaxy correlations can be adequately
described with $\xi(r,z) = (r_0/r)^\gamma (1+z)^{-(3+\epsilon)}$,
where the lengths are measured in proper units. This double power law
model does not allow any variation of the correlation slope, $\gamma$,
with redshift. The model might seem theoretically na\"ive, but it is
usually a better description of the data than any available non-linear
realization of a range of CDM models (\cite{ccc,virgo}). At low
redshift $\gamma=1.8$ is the canonical empirical value (\cite{dp}).
As redshift increases there are three very general possibilities as,
verified in n-body simulations (\cite{ccc}):
\par~~~$\bullet$ $\epsilon\simeq 0$ for particle clustering  if $\Omega_M$ low,
\par~~~$\bullet$ $\epsilon\simeq 1$ for particle clustering if  $\Omega_M\simeq 1$, and,
\par~~~$\bullet$ $\epsilon\simeq -1$ for dark matter halo clustering, with
        weak $\Omega_M$ dependence.
\par\noindent
The entire set of n-body particles defines the evolution of the mass
field. Halos with densities of $200\rho_c$ evolve in numbers as
described by the Press-Schechter model (\cite{ps}). However these
densities are too low to work as a one-to-one association of
galaxies inside groups and clusters, although they may be appropriate
for less clustered field galaxies. Halo cores at densities well above
the virialization density, $10^3-10^4\rho_c$, remain distinct even in
large clusters (\cite{cbg_cluster,moore}) and are the best candidates
for identification with galaxies in the context of purely
collisionless large scale structure simulations.

\begin{figure}
\centering
\vspace*{5pt}
\parbox{\textwidth}{\epsfysize=8truecm\epsfbox{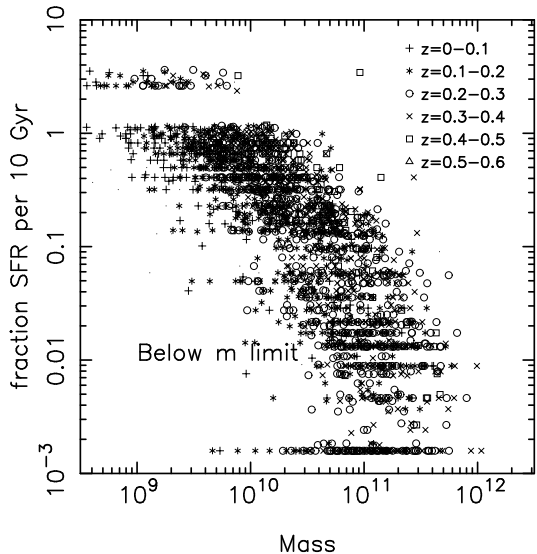}}
\vspace*{5pt}
\caption{The BC96 model estimated of the specific star formation
rate (multiplied with 15Gyr, assuming a duty cycle of unity) versus
the model mass for the CNOC2 galaxies. Galaxies of low mass and low
star formation rate will be below the $m_R$ limit of our survey.}
\label{fig:sfr_mass}
\end{figure}

A primary issue in studying the evolution of correlations is to be
able to identify the same population at different redshifts, since
both luminosity and colour dependence of clustering can mimic or mask
the desired effect. Although not yet fully implemented, fitting the
photometric SED to a model gives an indicative mass-to-light ratio
which can then be used estimate galaxy masses, as is shown in
Figure~\ref{fig:sfr_mass}. We use the Bruzual and Charlot (1996) SED
$\tau= $ 0.5, 1, 1.5, 2, 4 and 20~Gyr models and ages from 1 to 19 Gyr
in steps of 2 Gyr.  The very blue galaxies require 1~Gyr old models,
which is the cause of the gap in the indicative star formation rate of
Figure~\ref{fig:sfr_mass}. The fitted models give a stellar M/L and a
specific star formation rate, which is multiplied by 10~Gyr to give an
indication of the significance of star formation over a Hubble time,
assuming that the duty cycle for star formation is 100\%.  There are
two indicative results from Figure~\ref{fig:sfr_mass} which will
motivate our sample choice for correlation evolution estimation.
First, the galaxy mass function shows no significant evolution over
our redshift range, although a 30\% or so variation would be within
the limits.  Second, there is a clear confirmation of the strong
statistical correlation between star formation and the stellar mass of
galaxies.

High luminosity galaxies, $M_R^{k,e}\le -20$, are the closer than
lower luminosity galaxies to being a certifiable mass invariant
population, although the bluest members of this set likely have
significantly lower stellar masses which may not be well conserved
over our redshift range, if their star formation is steady in
time. This population has the considerable advantage that an
identically defined sample can be found in the LCRS.  The high
luminosity galaxies with $M_R^{k,e}\le -20$ comprise a volume limited
sample within the CNOC2 data, and approximate one within the LCRS
sample.

\begin{figure}
\centering
\vspace*{5pt}
\parbox{\textwidth}{\epsfysize=5truecm\epsfbox{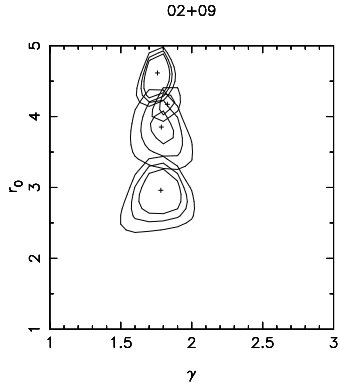}}
\vspace*{5pt}
\caption{Estimates of $r_0$ and $\gamma$ for high luminosity
galaxies from the LCRS and CNOC2 samples. The mean redshift, from top
to bottom is 0.080, 0.135, 0.28, and 0.43. Contours are 68, 90 and
99\% confidence levels.}
\label{fig:xiev_hi}
\end{figure}

\begin{figure}
\centering
\vspace*{5pt}
\parbox{\textwidth}{\epsfysize=5truecm\epsfbox{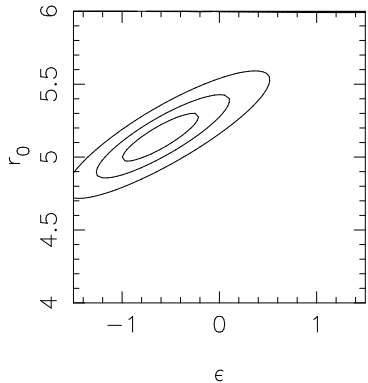}}
\vspace*{5pt}
\caption{Estimates of $r_0$ at $z=0$ and $\epsilon$ for high luminosity
galaxies from the LCRS and CNOC2 samples. The result falls between the
$\epsilon=0$ (fixed physical clustering) and $\epsilon=-1.2$ of clustering
fixed in co-moving co-ordinates.}
\label{fig:r0_eps}
\end{figure}

Two point correlation parameters, $r_0$ and $\epsilon$, for a
power-law correlation function of the high luminosity galaxies are
shown for various redshifts in Figure~\ref{fig:xiev_hi}. This redshift
sequence is fit to correlation evolution model,
$r_0(z)^\gamma=r_0^\gamma(1+z)^{-(3+\epsilon)}$, to estimate
$\epsilon$ and $r_0$, giving the result shown in
Figure~\ref{fig:r0_eps}. We find that $r_0=5.15\pm0.15\hmpc$ and
$\epsilon=-0.6\pm0.4$.  

These data strongly exclude clustering evolution that declines as
rapidly as $\epsilon=1$.  One could erroneously infer such a rapid
decline if one used a sample in which galaxies at higher redshifts are
intrinsically less luminous or on the average bluer than those nearby,
as is the case for the general population as a function of
redshift. The rate of correlation evolution falls between the values
expected for low $\Omega_M$ particles (non-merging) and dark matter
halos (which do merge).  Our result can be re-stated in terms of the
evolution of the correlation length in comoving co-ordinates, $x_0 =
r_0(1+z)^{-(3+\epsilon-\gamma)/\gamma)}$, as $x_0\propto
(1+z)^{-0.3\pm0.2}$.

It should be borne in mind that our estimate of correlation evolution
is a preliminary result and that the errors will be reduced with the
full sample.  An important consideration is that the luminosity cut
that defines our sample mixes together galaxies of a fairly wide range
of masses and anti-correlated star formation rates. It seems likely
that low and high star formation rate galaxies have different
correlation histories, which are mixed in the current approach.

\section{Galaxy Pairwise Velocities and Their Evolution}

The redshift space distortions in $\xi(r_p,r_z)$ reflect the dynamics
of clustering. The random velocities elongate contours of
$\xi(r_p,r_z)$ at small $r_p$ and infall velocities squash the
contours at large $r_p$. These distortions depend on $\Omega$, and the
biasing of galaxies with respect to the mass field.  The data here
have a local velocity accuracy of better than 100 \kms, as explicitly
demonstrated using redundant spectra taken through different
spectrograph slit masks. This is comparable to the velocity accuracy
of surveys at low redshift. The CNOC2 survey is designed to
concentrate on scales less than 10\hmpc\ so will not provide strong
limits on the infall velocities.

\begin{figure}
\centering
\vspace*{5pt}
\parbox{\textwidth}{\epsfysize=4truecm\epsfbox{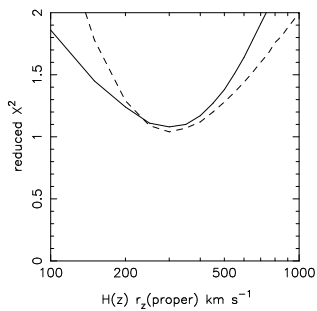}}
\vspace*{5pt}
\caption{The $\chi^2$ versus model
pairwise peculiar velocity, $\sigma_{12}$ within the LCRS sample
(solid line) at a mean redshift 0.10 and the CNOC2 sample (dashed
line) at a mean redshift of 0.36.  }
\label{fig:sig12_LH}
\end{figure}

The pairwise peculiar velocities are derived from a model
for $\xi(r_p,r_z)$ following the procedures of Croft, Dalton
\& Efstathiou (1998). Once the $r_0$ and $\gamma$ are derived
from the velocity independent correlations, then we set $\Omega_M=0.2$
and compute $\chi^2$ as a function of $\sigma_{12}$.  In
Figure~\ref{fig:sig12_LH} we show the reduced $\chi^2$ versus the
pairwise peculiar velocities for the LCRS sample at a mean redshift of
0.10 and the CNOC2 sample at mean redshift 0.36. The minimum of
$\chi^2$ rises slightly with redshift, although it is consistent with
no change with redshift. The $\epsilon$ model for clustering predicts
with the Cosmic Virial Theorem (see below) that $\sigma_{12}(z)\propto
(1+z)^{-\epsilon/2}$, provided that the bias is not changing with
redshift. We conclude that the peculiar velocities evolve in accord
with that predicted from the correlation function alone. This is
strong evidence that biasing of galaxies with respect to dark matter
is not a large effect at these redshifts.

\begin{figure}
\centering
\vspace*{5pt}
\parbox{\textwidth}{\epsfysize=4truecm\epsfbox{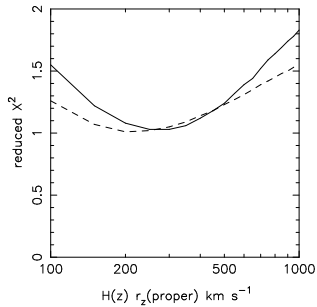}}
\vspace*{5pt}
\caption{The $\chi^2$ versus model pairwise peculiar velocity, $\sigma_{12}$
for the red (solid line) and blue galaxies (dashed line) within the
CNOC2 sample. There is between 70 and 100 \kms\ added in quadrature
with the true velocities.}
\label{fig:sig12_BR}
\end{figure}

The peculiar velocities also show a strong population dependence, as
shown in Figure~\ref{fig:sig12_BR}. The red galaxies have a pairwise
dispersion of about 350\kms, whereas the blue galaxies have a measured
dispersion of about 200\kms.  Removing the velocity errors in
quadrature would reduce the pairwise velocities of blue galaxies to
about 150\kms, which is a remarkably cold population. One can
speculate that in as much as accretion is necessary to promote ongoing
star formation, then star formation should be suppressed in regions of
strong tidal fields (the groups) and promoted in moderately dense
regions of low velocity dispersion.  A plot of the estimated star
formation rate versus local phase density indicates a weak, but
suggestive correlation.

The cosmic virial theorem (CVT) estimates the value of $\Omega_M$ in
the field (\cite{dp}) and is an important complement to studies of
clusters as $\Omega_M$ indicators. For a power law $\xi(r)$, the CVT
reads
$$
Q {\Omega_M\over b} = 
        {\sigma_{12}^2\over{(1+z)^3r_0^\gamma r^{2-\gamma} H_0^2}}
        {{4(\gamma-1)(2-\gamma)(4-\gamma)}\over{3J(\gamma)}},
$$
(\cite{dp}) where $Q$ is the three point correlation parameter, $b$
the linear bias of galaxy clustering relative to mass clustering, and
$J(\gamma)$ is 4.14 for $\gamma=1.7$ (\cite{lss}). For
$\sigma_{12}=350\pm50$, $r_0=3.2\pm0.2$ (the mean values over the
$0.15\le z \le 0.55$ range) we find $Q\Omega_M/b=0.11\pm0.04$. For
$Q=0.7$ (\cite{dp,bean}) this indicates $\Omega_M/b\simeq
0.15\pm0.06$. This is in good accord with the results from clusters
(\cite{glo,pro}) and the galaxy groups within this survey, neither of
which is significantly affected by bias. Because our sample finds both
a uniform luminosity dependence of clustering and a scale dependent
colour effect, all galaxies cannot have $b=1$. However, it is clear
that on 10\hmpc\ scales and smaller, that values of $b>2$ or $b<1/2$
are very unlikely. Moreover, this is further evidence that the matter
density of the universe is $\Omega_M\simeq 0.2$ and possibly somewhat
lower.

\section{Galaxy Groups}

\begin{figure}
\centering
\vspace*{5pt}
\parbox{\textwidth}{\epsfysize=5truecm\epsfbox{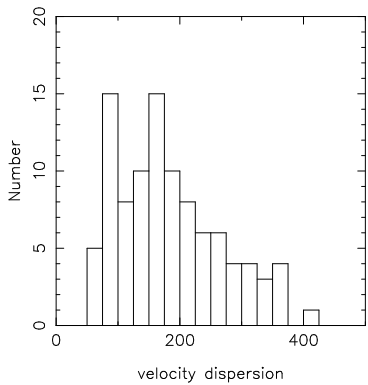}}
\vspace*{5pt}
\caption{The number of groups found as a function of their
line of sight (rest frame) velocity dispersion. The groups are
selected from a volume of about $2\times 10^5\Mpccube$.  Below 100
\kms\ there are few groups due to our velocity measurement error.}
\label{fig:nsig}
\end{figure}

Galaxy groups are the dense, quasi-virialized, regions within the
field. As virialized regions they offer several interesting
possibilities:
\par~~~$\bullet$ they can be used to estimate $\Omega_M$ via Oort's method,
\par~~~$\bullet$ the mean mass and luminosity profiles can be independently
	measured, and,
\par~~~$\bullet$ the relation between clustering and galaxy properties
        can be investigated.
\par\noindent
The highly nontrivial problem is to identify groups using redshift
space data. We adopt a simple algorithm. We ratio the number of
(selection function weighted) galaxies within a ``window'' to those in
a random sample, as for the correlation function, to measure the
redshift space density at the location of every galaxy. The window is
generally 0.5\hmpc\ in radius and $\pm600$ \kms\ in the redshift space
direction. We then select from this list the galaxy with the highest
overdensity and join to it all galaxies within a second, slightly
larger window, 1200 \kms\ and 0.8 \hmpc.  The sole output from this
operation is an estimate of the location of the centre of the group in
RA, Dec and z. The resulting centre has quite a weak dependence on the
details of the windows provided they are not so small that groups are
overlooked, or, that they are so big that huge fluffy regions are
selected as groups. The velocity cutoff is about three times the
pairwise velocity dispersion. The cutoff in projected radius is
20-30\% of the correlation length, approximately the radius at which
the redshift space correlation function, $\xi(s)$, bends due to random
velocities.  Our estimated centre is then given to an algorithm which
estimates the velocity dispersion, mass and total luminosity from
galaxies within some aperture, usually 0.4\hmpc\ and 900 \kms. These
values are not particularly crucial, since neither the projected
velocity dispersion nor the inferred $M/L$ has a measurable radial
gradient.

\begin{figure}
\centering
\vspace*{5pt}
\parbox{\textwidth}{\epsfysize=5truecm\epsfbox{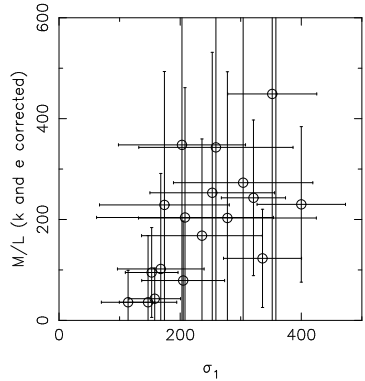}}
\vspace*{5pt}
\caption{The group mass-to-light ratio as a function of
velocity dispersion for groups with six or more members. The drop in M/L
at small velocity dispersion may be a result of correlated errors in
both axes, however it could have an origin in dynamical friction.}
\label{fig:ml_sig}
\end{figure}

\begin{figure}
\centering
\vspace*{5pt}
\parbox{\textwidth}{\epsfysize=5truecm\epsfbox{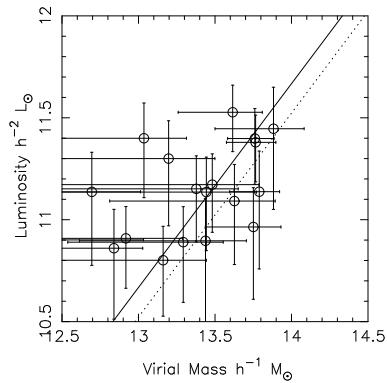}}
\vspace*{5pt}
\caption{Group luminosity plotted against mass. These quantities
are not intrinsically correlated by the measurement. The solid line is
the average group $M/L=215h\msun/\lsun$ and the dashed line is for the
population adjusted rich cluster $M/L=300h\msun/\lsun$. The more
massive groups appear to be consistent with clusters, but the smaller
ones appear to have lower M/L values. }
\label{fig:m_l}
\end{figure}

The groups are shown in Figure~\ref{fig:nsig} as a function of their
line of sight velocity dispersion. Although these groups should have
only a small population bias to estimate the global $M/L$, but they
need to be calibrated against n-body data if their cosmological number
densities are to be compared to Press-Schechter predictions. The
comforting news from this figure is that there is no evidence for many
high velocity dispersion groups which are likely to be bogus. If
anything, it is likely that the $\sim10\%$ of galaxies assigned to
groups here means that we are missing some genuine groups.

The virial mass-to-light ratio of the groups with six or more members
is shown as a function of velocity dispersion in
Figure~\ref{fig:ml_sig}.  The groups as a whole have a mean
mass-to-light of about $215h
\msun/\lsun$, and the groups as individuals appear to be nearly
consistent with having the same value. On the other hand, there is an
indication that lower velocity dispersion groups have lower $M/L$
values. The two axes in Figure~\ref{fig:ml_sig} are statistically
correlated, therefore we plot the mathematically independent
quantities of group luminosity versus virial mass in
Figure~\ref{fig:m_l}. The more massive (or equivalently, higher
velocity dispersion) groups have $M/L$ values consistent with the mean
value of $300h\msun/\lsun$ for CNOC1 clusters, after adjusting 0.12
mag to account for population differences and the slightly different
luminosity scales. An indication that lower luminosity groups are
``under-massive'' is still apparent in Figure~\ref{fig:m_l}. This is
perhaps not too surprising because there must be a scale at which the
$M/L$ characteristic of the mean field, $300h\msun/\lsun$ or so, drops
into the ``linear rise'' regime characteristic of individual isolated
galaxies (\cite{zw}).

\section{Internal Properties of Groups}

The cosmic virial theorem uses pairwise statistics, so its $\Omega_M$
is dependent on a knowledge of the bias of the tracer galaxies
relative to the mass field. On the other hand, if the single particle
velocity distribution can be found it allows the mass field to be
derived independently of the distribution of the tracer
galaxies. Galaxy groups provided that opportunity, since the
velocities can be measured with respect to the group mean
velocity. This means that we can put limits on the bias of galaxies
with respect to the mass field on scales of about 0.5\hmpc\ and less.

We use the groups whose local centers indicated velocity dispersions
more than 150 \kms\ and had six or more members. The following analysis
uses the group mean center (where there is usually no galaxy so the
density reaches a plateau) and includes all galaxies within 600 \kms\
in the frame of the group.  Jeans Equation allows the mass
field, $M(R)=\int\rho(r)\,dV$, to derived from a tracer population,
$\nu(r)$, which must be effectively in equilibrium, but not
necessarily distributed like the mass,
$$
M(r) = -{\sigma_r^2r\over G} \left[{{d \ln{\sigma_r^2}}\over{d\ln{r}}}
        + {{d\ln{\nu}}\over{{d\ln{r}}}} +2\beta\right].
$$
One complication is the velocity anisotropy parameter, $\beta= 1
-\sigma_\theta^2/\sigma_r^2$, which has a large effect near the center
but this uncertainty diminishes to about 20\% near the virial radius
(\cite{pro}).

\begin{figure}
\centering
\vspace*{5pt}
\parbox{\textwidth}{\epsfysize=5truecm\epsfbox{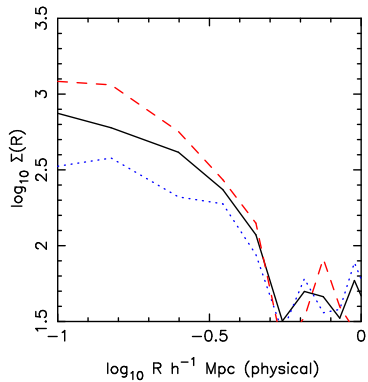}}
\vspace*{5pt}
\caption{The mean surface density of the groups (arbitrary units)
versus radius. The ratio of red (dashed) to blue (dotted) galaxies
increases towards the centre. The solid line is for all galaxies.}
\label{fig:Surf_R}
\end{figure}

\begin{figure}
\centering
\vspace*{5pt}
\parbox{\textwidth}{\epsfysize=5truecm\epsfbox{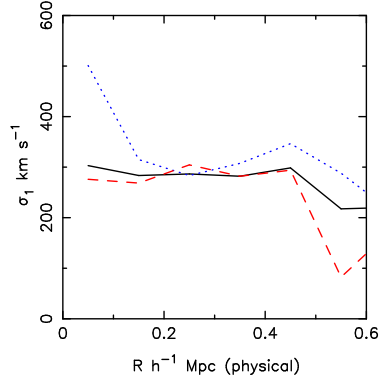}}
\vspace*{5pt}
\caption{The projected velocity dispersion versus radius
in the groups. Blue galaxies (dotted) have systematically higher
velocity dispersions than the red galaxies (dashed), which are similar
to the full sample (solid).  }
\label{fig:sig_R}
\end{figure}

The average group surface density profile, $\Sigma(R)$,
Figure~\ref{fig:Surf_R}, is based on about 200 galaxies.  The mean
projected slope between $\log{R}=-1$ and $\log{R}=-0.5$ of all the
galaxies is almost exactly $d\log(\Sigma)/d\log(r) = -1$, which
deprojects to $\nu\propto r^{-2}$ of the isothermal sphere.  The red
galaxies are more concentrated than the blue galaxies in these groups,
with small differences in the luminosity function from the field.  The
projected velocity dispersion profile, Figure~\ref{fig:sig_R}, of both
the full sample is consistent with no radial variation in the mass to
total light ratio. However the blue galaxies have significantly higher
velocity dispersions than the red galaxies with an indication of a
drop with radius. This is quite distinct from the pairwise velocities,
in which blue galaxies have significantly lower velocities than red
galaxies.  This helps one to understand why the virial masses given by
blue galaxies are about 50\% higher than the full sample.  Red
galaxies give virial masses that are 98\% of the full sample result.
We take these surface density and velocity dispersion profiles as
tentative evidence that the full galaxy distribution and the mass
distribution in groups are similar.  Within groups, blue galaxies are
``anti-biased'' with respect to the red galaxies and likely the mass
as well.

Although these results are both preliminary and clearly subject to
statistical uncertainties, they would support the notion that these
groups are in many ways dynamically similar to rich clusters although
they do not have such extreme population differences as rich clusters
and the field.  These issues are a major area of investigation in the
CNOC2 survey.

\section{Ultra-Large Scale Power}

Measuring the power spectrum on scales of 100-500\hmpc, roughly the
scale of the ``first Doppler peak'' in the CMB fluctuation spectrum,
requires a survey that covers an appreciable fraction of the visible
universe. Pencil beam surveys do just that, but only in one dimension
which leads to the problem of aliasing in of shorter scale power into
the derived 1D power spectrum, $P_1(k)$. In view of previous results
(\cite{beks}) and the fundamental interest in this statistic we
measure the one dimensional power spectrum from the CNOC2 data and
compare it to various simple models.

\begin{figure}
\centering
\vspace*{5pt}
\parbox{\textwidth}{\epsfysize=6truecm\epsfbox{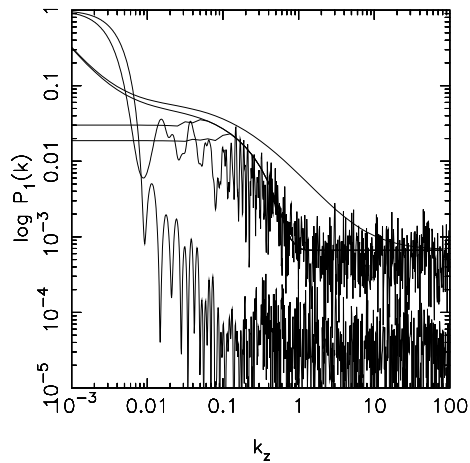}}
\vspace*{5pt}
\caption{The one dimensional power spectrum  and related
quantities. The lower irregular line is the power spectrum of the
smooth $n(z)$ distribution. The upper irregular line is the power
spectrum of the observed redshifts averaged over the two fields. The
upper smooth line is for $(r_0/r)^{1.7}$, with $r_0=4.2\hmpc$
(co-moving) which is modified with a Gaussian peculiar velocity
distribution of 300 \kms\ and then the power correlation truncated at
50 and 100
\hmpc.  }
\label{fig:pk1d}
\end{figure}

The one dimensional power spectrum $P_1(k) = |N^{-1}\sum_{i=1}^N
\exp{[ikx_i]}|^2$ (unweighted) is shown as the upper irregular line
in Figure~\ref{fig:pk1d}. At very large $k_z$ this tends to $1/N$
where $N\simeq 1500$ in each of our patches.  The observed 1D spectrum
requires some care in interpretation since the 1D power at a large
wavenumber is an integral over the entire 3D spectrum (\cite{kp})
whose analysis we follow here. The 1D power spectrum of the smooth
$n(z)$ is the $W(k_z)$ window function and is shown as the lower
irregular line. Clearly the power at $k_z<0.01 \hmpc^{-1}$ is
completely dominated by the smooth distribution of the data. At
shorter wavelengths, $\lambda<600\hmpc$, the observed power spectrum
is a result of real structure.

The null hypothesis is that the 1D power spectrum is simply the power
spectrum of the real space correlation function, $(r_0/r)^\gamma$, as
projected into these observations. The 3D power spectrum is $P(k) =
4\pi k^{-3}(kr_0)^\gamma\Gamma(2-\gamma)\sin{[(2-\gamma)\pi/2]}$.  To
derive the window function on the sky we do a Monte Carlo evaluation
of $W(k_x,k_y)$ by randomly populating the sky patches and
transforming. To evaluate the predicted 1D spectrum we approximate the
$z$ window function as $2\pi/L_z$, and integrate the product of the
sky window function with $P(k)$. The results are displayed in
Figure~\ref{fig:pk1d}, where we have added a constant shot noise of
$1/N$. The unaltered spectrum does a very poor job of describing the
observed spectrum. The two necessary modifications are to add a random
peculiar velocity and a long wavelength decline in power. The random
velocities are modeled using a Gaussian (clearly not a very good
description of these data) at 300 \kms.  Any CDM type spectrum rolls
over in the 30-100 \hmpc\ range.  We simply truncate the input power
spectrum at 50 and 100 \hmpc\ to produce the plots displayed. We can
conclude that the universe is not a fractal with a fixed clustering
dimension to all scales, and is otherwise consistent (at the current
statistical level) with the known large scale clustering power
spectrum (\cite{huanpk}).

\section{Preliminary Conclusions}

The CNOC2 sample spans a range of redshifts which has dramatic galaxy
evolution but in which detailed studies are readily performed. Our
preliminary conclusions are discussed below.

Galaxy evolution is confirmed to be strongly differential with the
blue, low mass galaxies changing significantly in their visible
numbers. We find luminosity functions similar in their overall
features to previous studies (\cite{cfrslf,ldss}) but with smaller
statistical errors.  The evolution of the early and intermediate SED
types is well described as a pure luminosity evolution. In contrast,
the late type SED population is described as nearly pure density
evolution. It should be noted that the density evolution could be at
least partly in the form of varying duty cycle of star formation.

The correlation of galaxies above $M_\ast+0.3$ mag is about 60\%
higher than those below, with a mean sample luminosity difference of
1.4 mag. This is consistent with the expected intrinsic correlation
differences of dark matter halos of different mass scales.  Having a
cleaner mass cut (from the SED model fits) and more data to trace the
evolution of this effect will allow a more confident conclusion.

Blue galaxies are about 50\% less correlated than red galaxies on
scales beyond about 0.3\hmpc. On shorter scales blue galaxies have a
break in their correlation function and become about a factor of three
less correlated than red galaxies. On large scales the difference is
about what one would expect from the smaller average masses of the
blue galaxies. Because the $\xi(r)$ break occurs at about the average
virialization scale it suggests that the blue-red difference is an
active environmental effect. Dynamically virialization occurs where
tidal stripping begins to remove the dark halos and their contents from
infalling galaxies.

\begin{figure}
\centering
\vspace*{5pt}
\parbox{\textwidth}{\epsfysize=5truecm\epsfbox{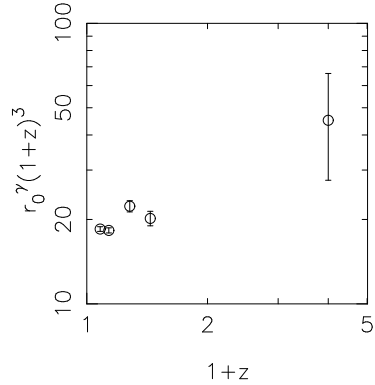}}
\vspace*{5pt}
\caption{The evolution of the proper density of galaxies
around galaxies, with the $M_R^{k,e}$ LCRS and CNOC2 sample at low
redshift and the Giavalisco \et\ (1998) inversion at high redshift.  }
\label{fig:Xiev}
\end{figure}

A primary result is that we find that the correlation evolution of
galaxies brighter than $M_R=-20$, where $M_\ast=-20.3$ mag, are well
described with $\xi(r) = (r_0/r)^\gamma(1+z)^{-3-\epsilon}$, where
$r_0=5.15\pm0.15\hmpc$ (proper co-ordinates), $\gamma=1.77\pm0.05$,
and $\epsilon=-0.6\pm0.4$.  That is, the physical density of galaxies
around galaxies is declining slightly with redshift.
Figure~\ref{fig:Xiev} shows that the extrapolation of our rate of
clustering evolution passes smoothly through the inversion of the
angular correlation of the high luminosity $z\simeq 3$ ``Steidel
objects'' (\cite{mauro}, but note \cite{adelberger}). We note that the
association of our population of objects with theirs is not secure.

The measured $\epsilon$ is completely incompatible with the
$\epsilon\simeq 1$ expected for the mass field in an $\Omega_M=1$
universe. The result is about half way between mass clustering in a
low density universe and the clustering evolution of the relatively
high density cores of dark matter halos (\cite{ccc}).  In general an
$\epsilon<0$ implies that bias is changing with redshift, although in
a low $\Omega_M$ model this is a very slow function of redshift. An
interpretation of the slow drop in proper density, within the context
of n-body simulations, is that it is due to merging of galaxies in
high density regions. Taking the result at face value, this means that
about $35\pm20\%$ of all the high luminosity galaxies will have merged
since redshift one. If the time scale for merging is 1~Gyr, then the
merging fraction will be about $4\pm3\%$, on average, close to the
estimates of Patton \et\ (1997).

The pairwise velocity dispersions of the luminous galaxies do not
detectably evolve between $z=0.05$ and $z=0.4$, remaining constant at
about 350 \kms\ (for our chosen estimator). The Cosmic Virial Theorem
gives $Q\Omega_M/b=0.11\pm0.04$, where $Q$ is the three point
correlation parameter. The three point parameter will be estimated
from these data, with the expectation that $Q\simeq 1$ from low
redshift results.

Galaxy groups with line-of-sight velocity dispersions between 100 and
400 \kms\ are found, with numbers declining with increasing
$\sigma_1$, very roughly as expected from Press-Schechter theory. The
virial M/L values may drop at small $\sigma_1$ (perhaps due to
dynamical inspiral) with the higher velocity dispersion objects having
M/L values consistent with the population adjusted values of rich
clusters. The mean M/L is about 35\% below rich clusters. The groups
are consistent with the total galaxy light being distributed like the
mass. The groups have a weak population gradient, becoming redder from
the ``edge'' of the mean group toward its centre. We tentatively
associate the break in the $\xi(r)$ of blue galaxies with this
behaviour of groups.  The groups, averaged together, indicate
$\Omega_M=0.15\pm 0.05$.

The one-dimensional power spectrum of our pencil beams requires that
the the observed power law correlation at scales of less than 10\hmpc\
must have a turnover at about 100\hmpc, which is of course expected
in CDM-like theories and indicated by observations of large scale
clustering at low redshift.

We set out to test theories of galaxy and structure evolution by
measuring the evolution. In spite of the dramatic changes in the
numbers of lower luminosity, blue galaxies, the systems of higher
luminosity that dominate the stellar mass of the universe appear to be
nearly a static population, having only a small change in the density
of similar galaxies around them and little star formation over our
redshift range.

These preliminary analyses are based on 1/2 of the eventual data for
which we complete observations in May 1998. The results are likely to
change somewhat (especially as other estimators are implemented and
systematic errors detected and eliminated). Perhaps the primary
conclusion to be drawn is that the CNOC2 dataset will be a rich sample
for the study of cosmology, large scale structure and galaxy
evolution.

\end{document}